\documentclass{elsart}

\usepackage{natbib}
\usepackage{epsfig}
\usepackage{amssymb}

\newcommand{\myemail}{caochen@bao.ac.cn}
\newcommand{\degree}{$^{\circ}$}
\def\astrobj#1{#1}

\begin{document}

\begin{frontmatter}

\title{Large Misalignment between Stellar Bar and Dust Pattern in \astrobj{NGC 3488} Revealed by {\it Spitzer} and SDSS}

\author[1,2]{C. Cao\corauthref{cor1}}
\ead{\myemail}
\author[1]{H. Wu} 
\ead{hwu@bao.ac.cn}
\author[3]{Z. Wang} 
\author[4]{L. C. Ho} 
\author[3]{J. S. Huang}
\author[5,1]{Z. G. Deng}
\address[1]{National Astronomical Observatories, Chinese Academy of Sciences, 
Beijing 100012, P. R. China}
\address[2]{Graduate School of the Chinese Academy of Sciences, Beijing 100039, China}
\address[3]{Harvard-Smithsonian Center for Astrophysics, 60 Garden Street, Cambridge, MA 02138}
\address[4]{The Observatories of the Carnegie Institution of Washington, 813 Santa Barbara 
Street, Pasadena, CA 91101}
\address[5]{College of Physical Sciences, Graduate School, Chinese Academy of Sciences, 
P.O. Box 3908, 100039 Beijing, China}
\corauth[cor1]{Fax: 86-10-64873566}

\begin{abstract}
A large position angle misalignment between the stellar bar and the distribution of dust 
in the late-type barred spiral \astrobj{NGC 3488} was discovered, using mid-infrared images 
from the {\it Spitzer Space Telescope} and optical images from the Sloan Digital Sky Survey 
(SDSS). The angle between the bar and dust patterns was measured to be 25\degree$\pm$2\degree, 
larger than most of the misalignments found previously in barred systems based on H$\alpha$ 
or H {\sc i}/CO observations. The stellar bar is bright at optical and 3.6$\mu$m, while 
the dust pattern is more prominent in the 8$\mu$m band but also shows up in the SDSS 
{\sl u} and {\sl g}-band images, suggesting a rich interstellar medium environment harboring 
ongoing star formation. This angular misalignment is unlikely to have been caused by spontaneous 
bar formation. We suggest that the stellar bar and the dust pattern may have different 
formation histories, and that the large misalignment was triggered by a tidal interaction 
with a small companion. A statistical analysis of a large sample of nearby galaxies with 
archival {\it Spitzer} data indicates that bar structure such as that seen in \astrobj{NGC 3488} 
is quite rare in the local Universe.

\end{abstract}


\end{frontmatter}


\section{Introduction}
\label{Sec1}
Bar structure as a major non-axisymmetric feature on all scales is important in 
studying the morphology, mass and light distributions (e.g., Freeman 1996; Elmegreen 
\& Elmegreen 1985; Elmegreen 1996; Elmegreen et al. 1996; Eskridge et al. 2000; 
Men{\'e}ndez-Delmestre et al. 2007), star formation (e.g., Zurita et al. 2004; 
Knapen 2005; Ondrechen \& van der Hulst 1983; Regan et al. 1996; Sheth et al. 2000), 
gas dynamics (e.g., Kormendy 1983; Bettoni \& Galletta 1988; Sancisi et al. 1979; 
Benedict et al. 1996; Downes et al. 1996; Regan et al. 1999) and central activities 
(e.g., Ho et al. 1997b; Hawarden et al. 1986; Knapen et al. 2002; Sakamoto et al. 
1999; Martini et al. 2003; Sheth et al. 2005) of disk galaxies. Theoretical models, 
including N-body and hydrodynamic simulations, generally confirm that bar formation 
is spontaneous and ubiquitous in disk evolution (e.g., Athanassoula 1992; Sellwood 
\& Wilkinson 1993; Friedli \& Benz 1993, 1995; Athanassoula \& Bureau 1999). Because 
of the dissipative nature of the interstellar medium (ISM), the streaming motions of 
the molecular gas in and around bar regions can be different from the stellar orbits 
(Athanassoula 1992; Regan et al. 1999; Sheth et al. 2002). Due to the delayed star 
formation after the clouds have been triggered ($\sim$30Myr; Vogel et al. 1988), the 
locations of gas/dust in galaxies can often be offset from that of young stars (e.g., 
Sheth et al. 2002; Phillips 1996; Martin \& Friedli 1997). The molecular gas can be 
transported from galactic disk toward central region by the gravitational torques 
from bars (e.g., Sakamoto et al. 1999; Sheth et al. 2002, 2005), and the condensation 
of gas leads to subsequent circumnuclear star formation (e.g., Ho et al. 1997b; Knapen 
et al. 2002; Martini et al. 2003; Jogee et al. 2005; Fisher 2006).

Observationally, the gas/dust patterns can often be seen as dust lanes, atomic and 
molecular gas concentrations, or isophotes of H {\sc ii} regions with active star 
formation (Martin \& Friedli 1997; Sakamoto et al. 1999; Regan et al. 1999; Rand et 
al. 1999; Crosthwaite et al. 2000; Sheth et al. 2002, 2005). As predicted by theoretical 
models (Athanassoula 1992; Friedli \& Benz 1993, 1995), there is a small position angle 
misalignment between the gas/dust distribution and the stellar bar, usually of a few 
(and up to 10) degrees, in the sense that the former is {\it leading}. Kenney et al. 
(1991) found the gaseous pattern is offset from the major axis of the stellar distribution 
by 24\degree $\pm$6\degree\/ in M~101. Crosthwaite et al. (2000) found that the central 
gas distribution as indicated by H {\sc i} map leads the stellar bar by almost 10\degree\/ 
in the late-type galaxy IC~342. Similarly, Rozas et al. (2000) identified a large sample 
of H {\sc ii} regions in barred galaxy \astrobj{NGC 3359} and showed a position angle 
misalignment of a few degrees exists in H$\alpha$ and I-band images. They also pointed 
out that the $u$-band image of this galaxy shows a bar pattern more aligned with  H$\alpha$, 
further suggesting massive star formation ``at the leading edge of the bar''. Sheth et al. 
(2002) found offsets between molecular gas (CO) and star formation (traced by H$\alpha$) 
in bars of six nearby spirals, which were caused by the gas flow dependent star formation. 
Understanding the misalignment between stellar and gas/dust patterns and their formation 
scenarios is crucial for studying the ISM properties and star formation processes taking 
place in environments where gas dynamics are strongly perturbed (e.g., Regan et al. 1996; 
Martin \& Friedli 1997; Sheth et al. 2000; Zurita et al. 2004), and also offers a good 
opportunity to study dynamical properties and secular evolution of barred galaxies (e.g., 
Kormendy 1983; Benedict et al. 1996; Regan et al. 1999; Kormendy \& Kennicutt 2004; Sheth 
et al. 2005; Kormendy \& Fisher 2005; Fisher 2006; Regan et al. 2006).

The {\it Spitzer Space Telescope}'s (Werner et al. 2004) observations in the mid-infrared, 
with its higher sensitivity and better angular resolution than previous observations (e.g., 
{\it ISO}), provide a new opportunity to study both stellar and gas/dust structures in 
galaxies (e.g., Pahre et al 2004; Wang et al. 2004; Cao \& Wu 2007). In particular, the four 
Infrared Array Camera (IRAC; Fazio et al. 2004) bands from 3.6 to 8.0 $\mu$m probe both stellar 
continuum and warm dust emissions (of the so-called polycyclic aromatic hydrocarbon, or PAH, 
and dust continuum emissions) with identical spatial sampling, thus enabling a powerful probe 
to compare non-axisymmetric features such as bar structures involving gas/dust and stellar 
mass. Recently, {\it Spitzer} observations of nearby galaxies have demonstrated the importance 
of using mid-infrared images for studying galaxy secular evolution driven by bar instabilities 
(e.g., Fisher 2006; Regan et al. 2006).

In this paper, we present an analysis of data from {\it Spitzer} and SDSS of the late-type 
barred spiral galaxy \astrobj{NGC 3488}. Previous studies show that, with an estimated distance 
of 39.9 Mpc (at this distance, 1$''$ corresponds to $\sim$193 parsecs) and a total infrared 
luminosity of $L_{\rm TIR}$ $\approx$ 4.6$\times$10$^{9}$ $L_{\rm \odot}$ (Bell 2003), 
\astrobj{NGC 3488} [Hubble type SB(s)c] has a weak bar ($\sim$1.5 kpc), with spiral arms 
beginning at the bar's end but without an inner ring. This is consistent with the conventional 
view that bars in most late-type spirals are relatively weak (Erwin 2005; Men{\'e}ndez-Delmestre 
et al. 2007), and that weak bars tend to produce a SB(s) type response (in which the spiral 
arms begin at the ends of the bar; Kormendy \& Kennicutt 2004). The data reduction is presented 
in $\S$2, and results on the bar structures in \astrobj{NGC 3488} with multi-wavelengths analysis 
are described in $\S$3. Possible explanations of the large misalignment between the bar 
and dust patterns are discussed in $\S$4.

\section{Data Reduction}
\label{Sec2}
Broad-band infrared images of \astrobj{NGC 3488} were acquired with IRAC on board {\it Spitzer}. 
The Basic Calibrated Data (BCD) were part of the Lockman Hole field in the {\it Spitzer} Wide-field 
Infrared Extragalactic (SWIRE) Survey (Lonsdale et al. 2003). Following the preliminary data 
reduction by the {\it Spitzer} Science Center pipeline, images of each of the four IRAC bands 
(3.6, 4.5, 5.8 and 8 $\mu$m) were mosaicked, after pointing refinement, distortion correction 
and cosmic-ray removal (Fazio et al. 2004; Huang et al. 2004; Wu et al. 2005; Surace et al. 2005; 
Cao \& Wu 2007; Wen et al. 2007). The mosaicked images have pixel sizes of 0.6$''$ and angular 
resolutions (full width at half maximum, FWHM) of 1.9$''$, 2.0$''$, 1.9$''$ and 2.2$''$ for the 
four bands, respectively. The angular resolution of IRAC 8$\mu$m images ($\sim$2.2$''$) is 
significantly improved over that of pre-{\it Spitzer} data at similar wavelengths (e.g., 
$\sim$10$''$ for ISOCAM LW2 at 7$\mu$m; Roussell et al. 2001).

In order to derive the dust-only 8$\mu$m component (PAH and dust continuum emissions), we remove 
the stellar continuum from the IRAC 8$\mu$m image by subtracting a scaled IRAC 3.6$\mu$m image 
(assuming that the 3.6$\mu$m emission is entirely due to old stellar population): 
\begin{displaymath}
f_{\rm \nu}(8\mu m)_{\rm dust} = f_{\rm \nu}(8\mu m) - \eta_{\rm 8\mu m}f_{\rm \nu}(3.6\mu m),
\end{displaymath}
where the scaling factor $\eta_{\rm 8\mu m}$ = 0.232 was calculated based on {\sl Starburst99} 
synthesis model (Leitherer et al. 1999), assuming solar metallicity and a Salpeter initial mass 
function between 0.1 and 120 $M_{\rm \odot}$. This approach has been adopted in several previous 
works (e.g., Helou et al. 2004; Wu et al. 2005; Regan et al. 2006; Bendo et al. 2006) for studying 
dust emissions and the 7.7$\mu$m PAH feature based on broad-band measurements, and shown to be effective.

The five-band optical images ($u,g,r,i,z$) and the fiber spectrum for the central region ($3''$ 
diameter) of \astrobj{NGC 3488} were taken from the SDSS data archive (York et al. 2000; Stoughton et al. 
2002). The background in each band was subtracted by fitting a low-order Legendre polynomial to it, 
after masking out all bright sources (Zheng et al. 1999; Wu et al. 2002). Figure 1 shows the three-color 
image of \astrobj{NGC 3488} derived from the SDSS data archive. North is up, and east is to the left, 
as denoted by the crosshair.

\section{Results}
\label{Sec3}

\subsection{Multi-wavelength View of Bar Structures in \astrobj{NGC 3488}}
The four-band IRAC images of \astrobj{NGC 3488} are shown in Figure 2. We 
find that \astrobj{NGC 3488} has two bar-like patterns that are bright in 
either stellar emission (at 3.6 and 4.5$\mu$m) or warm dust emission (at 8$\mu$m). 
We took the conventional approach of treating the bar as an elliptical feature 
for measuring its semi-major axis {\it a} and position angle PA. They can 
be measured with a fitting routine such as the {\tt ellipse} task in IRAF. 
The PAs are approximately 36\degree, 36\degree, 20\degree, 15\degree \/ for 
the bars in the 3.6, 4.5, 5.8, 8$\mu$m band, respectively. Uncertainties of 
the position angles are estimated to be $\pm 2$\degree, derived from the 
deviation of PAs along the major axis of the bar. Taking the mean of the two 
shorter wavelength bands as representing the stellar bar, we measured that it 
trails the dust pattern traced primarily by the continuum-subtracted 8$\mu$m 
emission (Fig. 3c), by a large position angle difference. Spiral arms with 
bright knots are also visible in the 8$\mu$m image: they appear to begin at 
the end of the dust pattern (Fig. 3d). 

For the SDSS images, the PA of the bar was also measured using {\tt ellipse}. 
The PA of the optical bar measured in this way is nearly identical ($\sim$40\degree) 
in the SDSS {\it g, r, i, z}-band images, and is also spatially coincident, 
within the measurement uncertainty of $\sim$$\pm 2$\degree, with that of the IRAC 
3.6 and 4.5$\mu$m images, but trails the dust pattern at 8$\mu$m by 25\degree\/ 
(Fig. 3b). However, the bar in the SDSS $u$-band is quite different from that in the 
other SDSS bands: at PA $\approx 15$\degree, it is instead much better aligned with 
the dust pattern bright at IRAC 8$\mu$m (Fig. 3a). 

The deprojected values of the relative length [L$_{\rm b}(i)$] and the misalignment 
between the stellar bar and the dust pattern [$\theta$(i)] are calculated using 
the equations given by Martin (1995):
\begin{displaymath}
L_{\rm b}(i) = \frac{2a[cos^{2}(\phi_{\rm a}) + sec^{2}(i)sin^{2}(\phi_{\rm a})]^{1/2}}{D_{\rm 25}}
\end{displaymath}
\begin{displaymath}
\theta(i) = arctan[cos(i)tan(\phi_{\rm a;8\mu m})] - arctan[cos(i)tan(\phi_{\rm a;stellar})],
\end{displaymath}
where $i$ is the inclination angle of \astrobj{NGC 3488} ($\sim$49.5\degree, calculated 
using the formula given by Bottinelli et al. 1983), $a$ is the length of the 
semi-major axis, $\phi_{\rm a}$ is the angle between the patterns (dust or stellar) 
and the node lines (the major axis defines the PA of the galaxy, $\sim$175\degree), 
and $D_{\rm 25}$ is the diameter of NGC~3488 at a $B$-band surface brightness of 25 mag 
arcsec$^{-2}$ ($\sim$1.86 arcmin).
\footnote{The PA, $D_{\rm 25}$, and $R_{\rm 25}$ of \astrobj{NGC 3488} were taken from 
The Third Reference Catalogue of Bright Galaxies (de Vaucouleurs et al. 1991).}
 These parameters of the patterns are summarized in Table 1. The angular misalignment 
between the optical bar and the infrared dust pattern, $\theta(i)$, is approximately 
20\degree \/ after correcting for inclination effect. Uncertainties of these parameters 
can be as large as $20\%$, due mostly to simplistic assumptions concerning projection 
effects and the difficulty in decoupling the bar from the bulge (Martin 1995; Martin \& 
Friedli 1997), especially in the IRAC 8$\mu$m image. Nevertheless, this angular misalignment, 
assuming it represents the misalignment between the dust and stellar patterns, belongs 
to one of the largest misalignments found in previous observations or in typical numerical 
simulations.

Besides the bright stellar bar, a faint and clumpy bar-like structure appears to be present 
in the SDSS $g$-band image (Fig. 4, left panel). We use two different techniques to enhance 
the visibility of this structure: unsharp masking (e.g., Walterbos et al. 1994) and 
deconvolution (Fig. 4, right panel). Unsharp masking was made by using a 2-pixel wide 
(2$\sigma$) Gaussian filter. The deconvolution was performed with the task {\tt lucy} 
(Lucy 1974) in IRAF, using a point-spread function derived from the associated psField 
file of the SDSS data; the total number of iterations is 30. The PA of this faint bar-like 
structure is approximately 15\degree $\pm$2\degree, so it is spatially coincident with the 
patterns in $u$-band and 8$\mu$m images.

\subsection{Stellar Populations in Bars}
The bar bright at optical (SDSS) and IRAC 3.6, 4.5 $\mu$m is known to be dominated by 
old stellar population. Gadotti \& de Souza (2006) showed that the bar color index can be 
used as an indicator of the bar age:  old bars are on average redder than young ones. The 
SDSS $g-r$ and $g-i$ colors of the stellar bar of \astrobj{NGC 3488} were compared with an 
instantaneous burst model calculated based on GALAXEV (Bruzual \& Charlot 2003), adopting 
a Salpeter initial mass function and solar metallicity as initial conditions. This comparison 
shows that the stellar bar bright in optical and IRAC 3.6 and 4.5$\mu$m bands is evolved, 
with a mean age of the order of 4 Gyr. 

The pattern bright at 8$\mu$m is dominated by the strong 7.7$\mu$m PAH feature and warm dust 
continuum emission from very small grains, while the pattern shown in the IRAC 5.8$\mu$m image 
is probably a mixture of stellar and dust (6.2$\mu$m PAH and warm dust) components. From the 
correlation between 8$\mu$m dust emission and star formation activities (Wu et al. 2005; 
Calzetti et al. 2005) and the previous result that 8$\mu$m dust and 24$\mu$m hot dust emission 
are well correlated on kiloparsec scales
\footnote{MIPS 24$\mu$m emission, which is mainly due to hot dust emission from very small grains, 
is thought to be a good measure of the SFR in galaxies (e.g., Wu et al. 2005; Calzetti et al. 2005; 
P{\'e}rez-Gonz{\'a}lez et al. 2006).}
 (e.g., Bendo et al. 2006), we suggest that the dust pattern seen at 8$\mu$m is associated with 
young stars and thus represents recent star forming activity. Some authors (e.g., Regan et al. 
2006; R. C. Kennicutt, priv. comm.), however, argue that PAH emission is likely a better tracer 
of the general ISM rather than star-forming regions heated by young, massive stars. The coincidence 
of the bar shown in the $u$-band with that bright at 8$\mu$m (Fig. 3a) confirms that there 
exists a young stellar population in the region of the dust pattern. 

\subsection{Star Formation Rates}
The star formation rate (SFR) in the bar region of \astrobj{NGC 3488} can be estimated using 
either the H$\alpha$ line flux or the 8$\mu$m dust emission. The SDSS spectrum, taken through a 
3$''$-diameter ($\sim$0.58 kpc) fiber, indicates that the central region can be classified as 
an ``H {\sc ii} nucleus'' (Ho et al. 1997a), one whose main source of ionizing photons derives 
from young stars. The H$\alpha$-based SFR is calculated using Equation B3 of Hopkins et al. 
(2003), after correcting for extinction using the observed Balmer decrement (Calzetti 2001):
\begin{displaymath}
SFR_{\rm H\alpha}(M_{\rm \odot}{\rm yr}^{-1}) = 4\pi D_{\rm l}^{2}S_{\rm H\alpha}(\frac{S_{\rm H\alpha}/
S_{\rm H\beta}}{2.86})^{2.114}\frac{1}{1.27\times10^{34}},
\end{displaymath} 
where $S_{\rm H\alpha}$ and $S_{\rm H\beta}$ are the stellar absorption-corrected line fluxes, 
derived from the emission-line catalog given by MPA-SDSS (Tremonti et al. 2004). The measured 
SFR for the central 3$''$ based on H$\alpha$ is 0.0077 $M_{\rm \odot}$ yr$^{-1}$.

To estimate the SFR along the bar, we use the 8$\mu$m dust emission, which is thought to be 
measure of the SFR in galaxies (Wu et al. 2005).  From Equation 4 of Wu et al. (2005):
\begin{displaymath}
SFR_{\rm 8\mu m\ dust}(M_{\rm \odot}{\rm yr}^{-1}) = \frac{\nu L_{\rm \nu}({\rm 8\mu m \ dust})}{1.57\times10^{9}L_{\rm \odot}}.
\end{displaymath} 
An aperture of 3$''$ diameter was selected for measuring the SFR centered on the nucleus using 
the dust-only 8$\mu$m image, to enable a sensible comparison with the SFR derived from the H$\alpha$ 
flux. The SFR measured in this way is 0.0079 $M_{\rm \odot}$ yr$^{-1}$, consistent with that derived from
H$\alpha$. This result indicates that the 8$\mu$m dust emission can be used as a tracer of the SFR in 
the central region of galaxies, at least for the case of \astrobj{NGC 3488}. The enhanced 8$\mu$m emission in 
the central region of \astrobj{NGC 3488} is consistent with the result that barred galaxies tend to have 
strong central excesses in 8$\mu$m emission (Regan et al. 2006), which suggests that bars induce gas inflows 
toward the center of galaxies as an internal process of galaxy secular evolution (e.g., Sakamoto et al. 1999; 
Sheth et al. 2000, 2002, 2005; Jogee et al. 2005). We also estimated the total SFR along the bar after excluding 
the contribution from the central region. The photometry was performed on the dust-only 8$\mu$m image using an 
ellipse with a semi-major axis of 3$''$ and an ellipticity of 0.5, chosen to match the isocontour of the dust 
pattern. The measured SFR for the bar and nucleus is $\sim$0.0150 $M_{\rm \odot}$ yr$^{-1}$, which implies that 
the SFR along the bar is roughly 0.0072 $M_{\rm \odot}$ yr$^{-1}$, comparable to the value at the nucleus.

\section{Discussion}
\label{Sec4}

\subsection{Possible Formation Scenarios of the Misalignment Between Bars in \astrobj{NGC 3488}}
Martin \& Friedli (1997) found that the H$\alpha$ bars (tracing young stars and H {\sc ii} 
regions) and stellar bars are misaligned by up to 10\degree \/ among 11 barred galaxies. 
Friedli \& Benz (1993, 1995), using their N-body + SPH simulations of spontaneous bar formation, 
showed that the H {\sc ii} regions tend to lead the stellar bar by an angle of several degrees. 
In these simulations the misalignments between bars are similar to those observed, and they 
occur as a result of orbit crossings of gas motion, mostly at early epochs during the formation 
of a strong, fast-rotating bar in the absence of an inner Lindblad resonance (Martin \& Friedli 1997). 
The best example is \astrobj{NGC 3359}, in which star formation is completely absent in the galaxy 
nucleus (Martin \& Friedli 1997), and the age of the bar is young ($\sim$400 Myr; Martin \& Roy 1995). 
But this is unlikely to be the case in \astrobj{NGC 3488}, due to the fact that the age of its 
stellar bar ($\sim$4 Gyr) is much older than bars in the spontaneous bar formation scenario 
(e.g., \astrobj{NGC 3359}).

Alternatively, perhaps the morphology of this galaxy represents an episode shortly after the 
capture of a small, secondary galaxy. In such a galaxy merger scenario, the stellar and dust 
patterns may have different formation histories. The stellar bar may have formed previously as 
in a spontaneous formation scenario, and the dust pattern could be tidally induced later from a 
dwarf galaxy swallowed by \astrobj{NGC 3488}. The different stellar populations in the stellar 
bar and the dust pattern (see \S3.2) also support this scenario. Berentzen et al. (2003) investigated 
the dynamical effects of the interaction between an initially barred galaxy and a small companion, 
using N-body/SPH numerical simulations. They found that the interactions can produce offset bars, 
nuclear and circumnuclear disks, and tidal arms connected to the end of the bar. Based on their 
results that the fate of the stellar bar is determined by the impact position, we speculate that 
in \astrobj{NGC 3488} there may have been an impact on the bar major axis when the bar was weak. 
In such a case, the tidal force exerted on the bar does not disrupt much the bar structure (i.e., the 
stellar bar survived after the impact). The nearby bright, compact object toward the northeast 
of the SDSS images (see Fig. 3b), identified as SDSS J110124.04+574045.1, is a plausible candidate 
for a dwarf galaxy that may have hit \astrobj{NGC 3488}. Its colors are very blue ($u-g = 0.309$, 
$g-r = -0.222$, $r-i = -0.757$) and similar to those of irregular galaxies (Fukugita et al. 1995). 
And its absolute magnitude ($-12.73$, $-13.59$, $-13.93$, $-13.82$, $-13.81$ for {\sl u, g, r, i, z}, 
respectively) are located at the faint end of the luminosity function of extremely low-luminosity 
galaxies (Blanton et al. 2005), if we assume that its distance is the same as that of \astrobj{NGC 3488} 
(39.9 Mpc). However, we cannot exclude the possibility that it is a foreground white dwarf star 
superposed on \astrobj{NGC 3488}, since its very blue colors are consistent with those of spectroscopically 
identified white dwarf stars in SDSS (Kleinman et al. 2004).  

\subsection{Frequency of Large Misalignments Between Bars Among Galaxies Revealed By {\it Spitzer}}
Large misalignments ($>$10\degree) between stellar bars (bright at optical and 3.6$\mu$m) and dust 
distributions (shown at 8$\mu$m) are quite rare among nearby barred galaxies with archival {\it Spitzer} 
data. We have examined $\sim$50 barred spirals (SB and SAB) in the {\it Spitzer} Infrared Nearby 
Galaxies Survey (SINGS, Kennicutt et al. 2003) and in the Mid-IR Hubble Atlas of Galaxies (a {\it Spitzer} 
GTO program, PID: \#69, PI: G. Fazio; see also Pahre et al. 2004), and found that {\it none} of them 
shows a misalignment as large as that found in \astrobj{NGC 3488}. Most of the galaxies only have a 
single old stellar bar, which is bright at IRAC 3.6 and 4.5 $\mu$m but absent at 8$\mu$m (e.g., 
\astrobj{NGC 7080}). Others that have younger bars always show good alignment between 3.6 and 8$\mu$m 
(e.g., \astrobj{NGC 7479}, consistent with the previous result that no misalignment was observed between 
its H$\alpha$ and stellar bars; Martin \& Friedli 1997). This statistical evidence indicates that the 
bar structure seen in \astrobj{NGC 3488} is quite rare among barred galaxies in the local Universe, 
and that the misalignment between bar and dust patterns may be a short-lived phenomenon in the evolutionary 
history of the galaxy. However, any firm conclusion must await a quantitative analysis of a large, 
well-defined, and unbiased sample of barred galaxies.

\section{Summary}
\label{Sec5}
Using mid-infrared images from {\it Spitzer} and optical images from SDSS, we show that the late-type 
barred spiral galaxy \astrobj{NGC 3488} contains two misaligned patterns, one composed of old stars 
and the other young stars and dust. The angle between the two patterns ($\sim$25\degree) is among the 
largest ever reported. The stellar bar is bright in the optical and in the IRAC 3.6 and 4.5$\mu$m 
bands, and is dominated by old stars. The dust pattern is more prominent in the 8$\mu$m band, but also 
shows up in the SDSS {\sl u} and {\sl g} bands; it traces regions of recent or ongoing star formation. 
The dust pattern could be tidally induced by a dwarf galaxy swallowed by \astrobj{NGC 3488}. From 
examination of mid-infrared images of a large sample of nearby barred galaxies with archival {\it Spitzer} 
data, we find that bar structure such as that found in \astrobj{NGC 3488} is quite rare in the local Universe.

To further test the hypothesis that the large misalignment in \astrobj{NGC 3488} was triggered by 
a recent merger, it would be desirable to obtain deeper imaging observations of the system in order 
to search for morphological features suggestive of tidal interactions. Obtaining a spectrum of 
the candidate dwarf galaxy would help validate its physical association with \astrobj{NGC 3488}. 
Additional numerical simulations will help to validate whether two large-scale patterns can coexist 
over long time scales, since it is possible that the competing torques will introduce chaos to the 
system (I. Berentzen, priv. comm.).

~\\
{\bf Acknowledgments}\\
~\\
We would like to thank the anonymous referee for very constructive comments and suggestions. 
We thank X.-Y. Xia, S. Mao, R. Kennicutt, I. Berentzen for advice and helpful discussions, 
and C.-N. Hao, J.-L. Wang, F.-S. Liu for their capable assistance throughout the 
process of {\it Spitzer} data reductions. This project is supported by NSFC No.10273012, 
No.10333060, No.10473013, No.10373008. This work is based in part on observations made with 
the {\it Spitzer Space Telescope}, which is operated by the Jet Propulsion Laboratory, 
California Institute of Technology under NASA contract 1407. Funding for the SDSS and 
SDSS-II has been provided by the Alfred P. Sloan Foundation, the Participating Institutions, 
the National Science Foundation, the U.S. Department of Energy, the National Aeronautics 
and Space Administration, the Japanese Monbukagakusho, the Max Planck Society, and the 
Higher Education Funding Council for England. The SDSS Web Site is http://www.sdss.org/. 
The SDSS is managed by the Astrophysical Research Consortium for the Participating Institutions. 
The Participating Institutions are the American Museum of Natural History, Astrophysical 
Institute Potsdam, University of Basel, Cambridge University, Case Western Reserve University, 
University of Chicago, Drexel University, Fermilab, the Institute for Advanced Study, the 
Japan Participation Group, Johns Hopkins University, the Joint Institute for Nuclear Astrophysics, 
the Kavli Institute for Particle Astrophysics and Cosmology, the Korean Scientist Group, the 
Chinese Academy of Sciences (LAMOST), Los Alamos National Laboratory, the Max-Planck-Institute for 
Astronomy (MPIA), the Max-Planck-Institute for Astrophysics (MPA), New Mexico State University, 
Ohio State University, University of Pittsburgh, University of Portsmouth, Princeton University, 
the United States Naval Observatory, and the University of Washington.

\clearpage

\begin{figure}
\center
\includegraphics[angle=0,scale=.8]{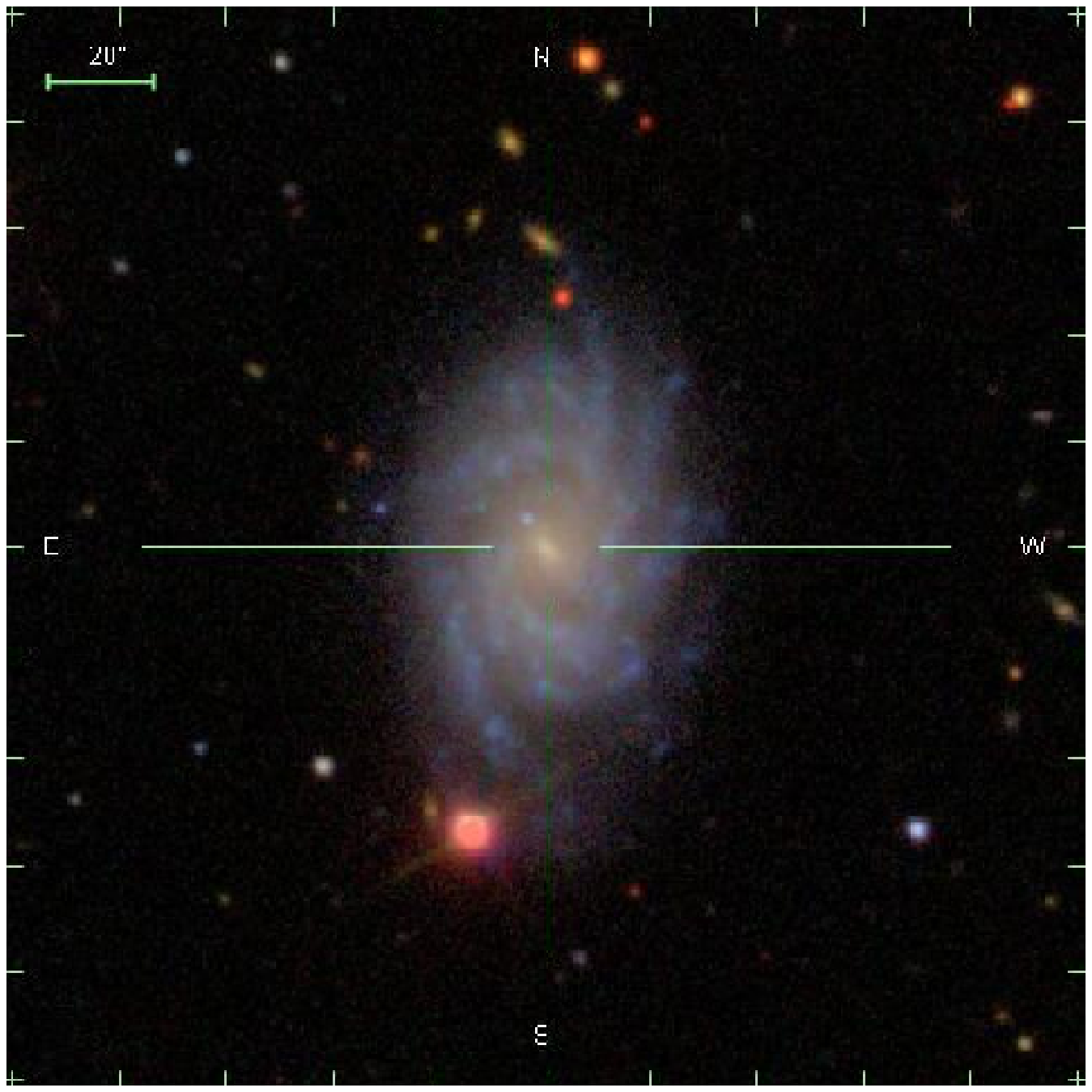}
\caption{Three-color image of \astrobj{NGC 3488} derived from the SDSS data archive. 
North is up, and east is to the left, as denoted by the crosshair. \label{fig1}}
\end{figure}

\clearpage

\begin{figure}
\center
\includegraphics[angle=0,scale=.7]{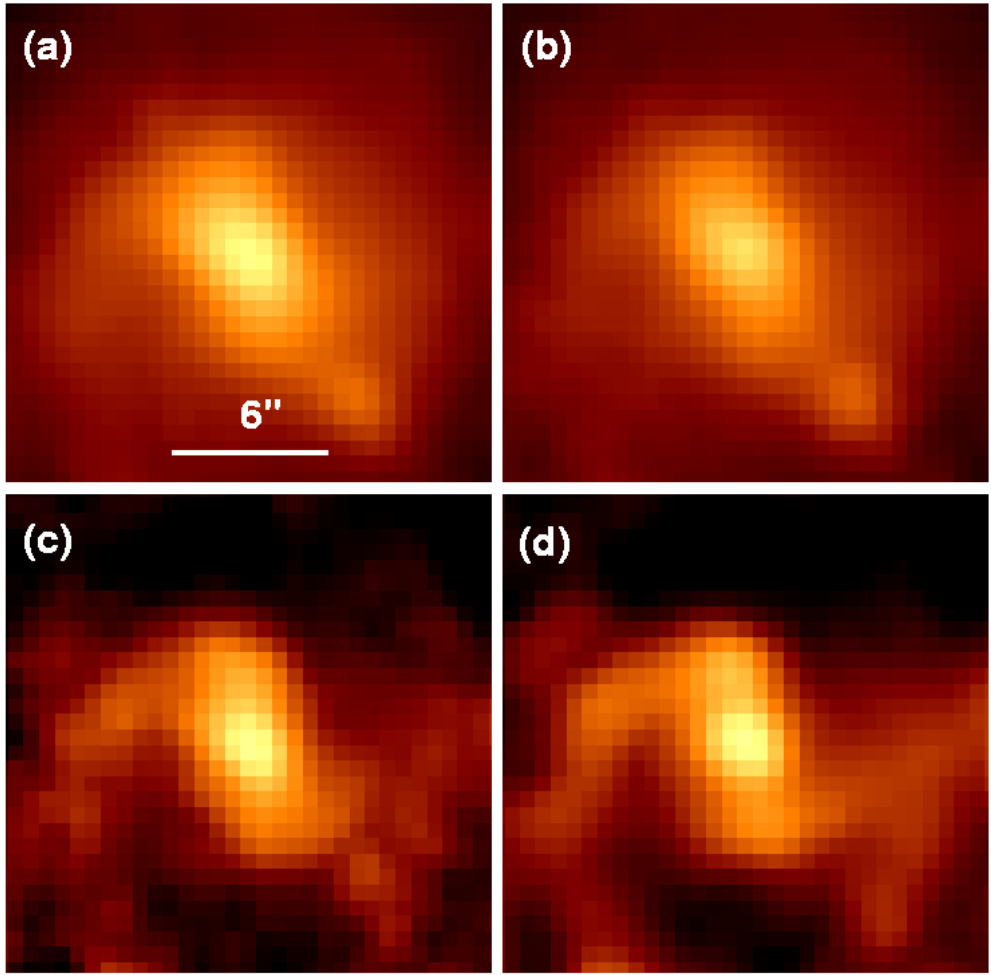}
\caption{The {\it Spitzer} IRAC four-band images [from (a) to (d): 3.6, 4.5, 5.8, 8$\mu$m] of the 
bar region in \astrobj{NGC 3488}. North is up; East is to the left. The stellar bar bright at 
3.6 and 4.5 $\mu$m and the dust pattern bright at 8 $\mu$m are misaligned by a large angle. \label{fig2}}
\end{figure}

\clearpage

\begin{figure}
\center
\includegraphics[angle=0,scale=.7]{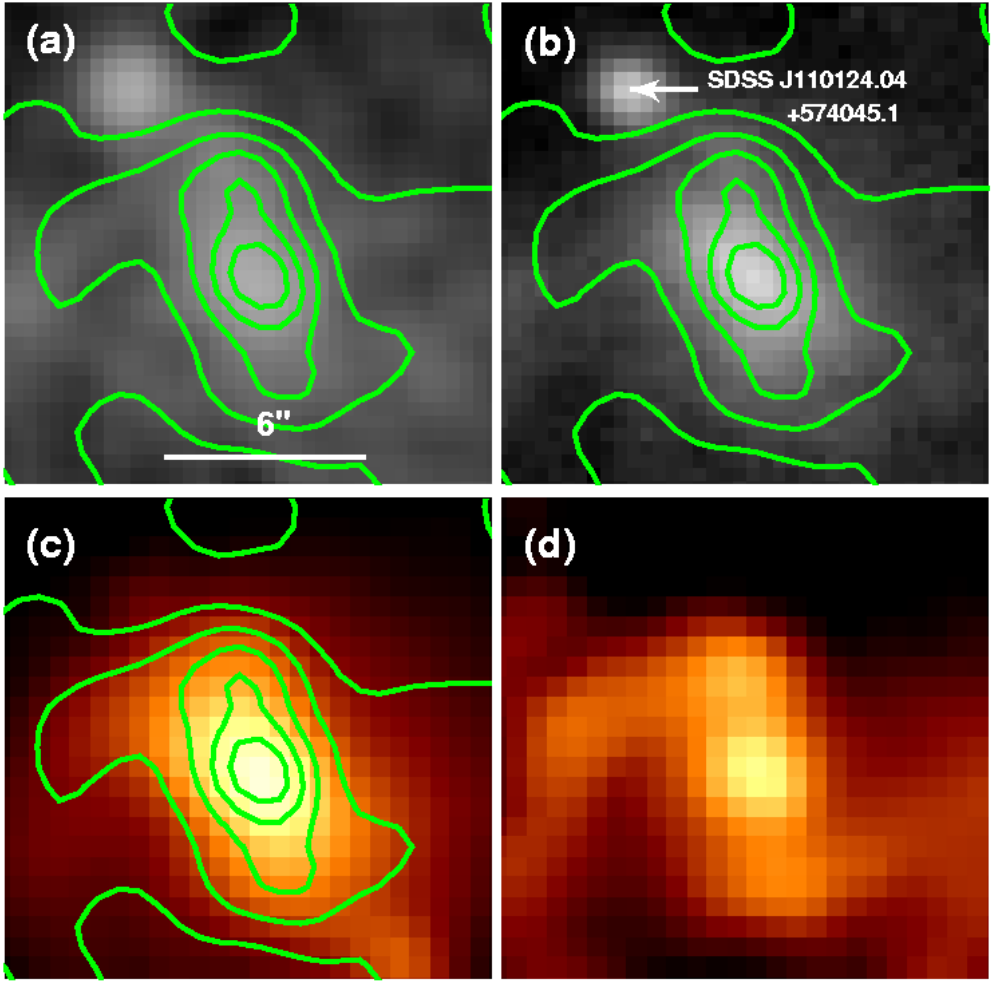}
\caption{From (a) to (d): the SDSS $u$-band, $g$-band, IRAC 3.6$\mu$m, and dust-only 8$\mu$m images of 
the bar region in \astrobj{NGC 3488}. North is up; East is to the left. The $u$-band image was smoothed 
using a 2-pixel wide (2$\sigma$) Gaussian filter. Contours of the dust-only 8$\mu$m image are superposed 
on the $u$-band, $g$-band, and IRAC 3.6$\mu$m images. A large misalignment by an apparent angle ($\theta$) 
of about 25\degree\/ is shown between the bright bar at optical (with a PA of $\sim$40\degree\/) and the 
dust distribution shown in the dust-only 8$\mu$m image (with a PA of $\sim$15\degree\/). A nearby compact, 
bright object (SDSS J110124.04+574045.1) is shown in the northeast of the SDSS images. \label{fig3}}
\end{figure}

\clearpage

\begin{figure}
\center
\includegraphics[angle=0,scale=.7]{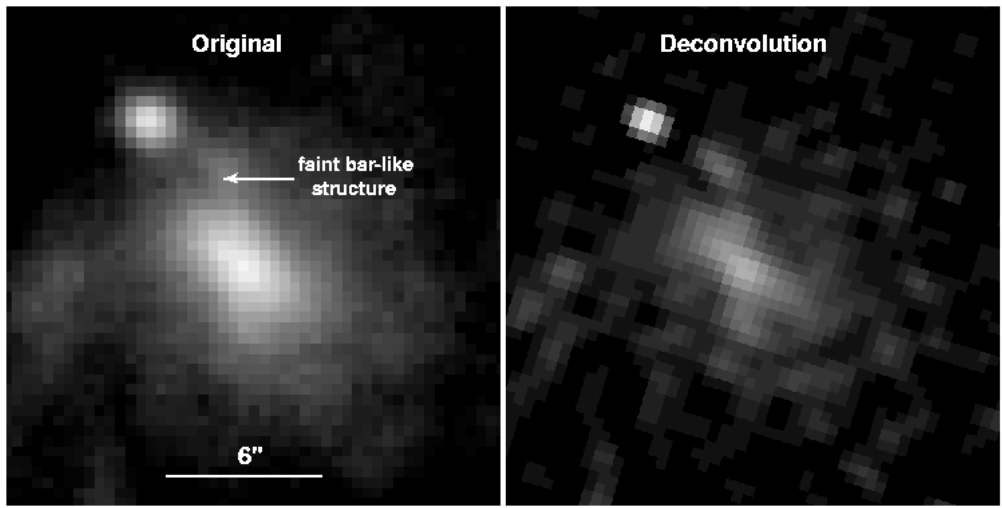}
\caption{The original (left) and deconvolved (right) SDSS $g$-band images of the bar region in \astrobj{NGC 3488}. 
North is up; East is to the left. A faint and clumpy bar-like structure is suggested in the deconvolved 
images.\label{fig4}}
\end{figure}

\clearpage

\begin{table}[]
  \begin{center}
  \caption[]{Properties of bar and dust patterns}
  \label{Tab:1}
  \begin{tabular}{ccccccc}
  \hline
  \hline
Regions & $a$[''] & $L_{\rm b}$(i) & PA[\degree] & $\theta$[\degree] & $\theta$(i)[\degree] & Image\\
~    & (1) & (2) & (3) & (4) & (5) & (6)\\
\hline
Stellar Bar  & 4.2 & 0.111 & 40.0 & ~ & ~ & SDSS $g$ band\\
~            & ~ & ~ & ~         & 25.0 & 19.7 & ~     \\
Dust Pattern     & 3.0 & 0.067 & 15.0 & ~ & ~ & IRAC 8$\mu$m\\
\hline
  \end{tabular}\end{center}
\begin{list}{}{}
\item[]Notes.$\--$Col.(1): The lengths of semi-major axis of the stellar bar and dust pattern; Col.(2): 
The relative lengths of the patterns; Col.(3): Position angles measured in the conventional manner, 
from North through East; Col.(4): The measured angular misalignment between stellar and dust patterns; 
Col.(5): The angular misalignment after correcting for inclination effect; Col.(6): Images used for the 
measurements.
\end{list}
\end{table}

\end{document}